\documentclass[journal]{IEEEtran}
\usepackage{amsfonts}
\IEEEoverridecommandlockouts

\ifCLASSINFOpdf
\else
\fi
\usepackage{epstopdf}
\usepackage{multirow}
\usepackage{cite}
\usepackage{stfloats}
\usepackage{color}
\usepackage{epsfig}
\usepackage{graphicx}
\usepackage{psfig}
\usepackage{epsf}
\usepackage{slashbox}
\usepackage{float}
\usepackage{amssymb }
\usepackage[cmex10]{amsmath}
\begin{document}
\title{Heterogeneous  Power-Splitting Based Two-Way DF Relaying with Non-Linear Energy Harvesting}
\author{\IEEEauthorblockN{Liqin Shi$^1$, Wenchi Cheng$^1$, Yinghui Ye$^1$, Hailin Zhang$^1$, and Rose Qingyang Hu$^2$}\\
\IEEEauthorblockA{$^1$State Key Laboratory of Integrated Services Networks, Xidian University, China\\
 $^2$Department of Electrical and Computer Engineering, Utah State University, USA}
}
\maketitle

\begin{abstract}
Simultaneous wireless information and power transfer (SWIPT) has been recognized as a promising approach to improving the performance of energy constrained networks.
In this paper, we investigate a SWIPT based three-step two-way decode-and-forward (DF) relay network with a  non-linear energy harvester equipped at the relay.
As most existing works require instantaneous channel state information (CSI) while CSI is not fully utilized when designing power splitting (PS) schemes, there exists an opportunity for enhancement by exploiting CSI for PS design.
To this end,
we propose a novel heterogeneous PS scheme, where the PS ratios are dynamically changed according to  instantaneous channel gains. In particular, we derive the closed-form expressions of the optimal PS ratios to maximize the capacity of the investigated network and analyze the outage probability with the optimal dynamic PS ratios based on the non-linear energy harvesting (EH) model. The results provide valuable insights into the
effect of various system parameters, such as  transmit power of the source, source transmission rate, and source to relay distance on the performance of the investigated network. The results show that our proposed  PS scheme outperforms the existing schemes.

\end{abstract}
\IEEEpeerreviewmaketitle
\section{Introduction}
Recently, simultaneous wireless information and power transfer (SWIPT) has emerged as an appealing approach to prolonging  the lifetime of energy-constrained networks, e.g., relay networks \cite{7744827,RandomPS}, wireless sensor networks \cite{8187650}, cooperative non-orthogonal multiple access networks
\cite{8108186}, D2D assisted cellular networks \cite{ShiTvt}, by harvesting energy from radio frequency (RF) signals.
Of particular interest is integrating SWIPT with relay networks, which not only extends the wireless transmission range, but also prolongs the operating time of the energy-constrained relay nodes \cite{R1,8108186}.
Compared with one-way relaying, two-way relaying, which can be performed in two steps or three steps, can offer a more efficient use of the available resources by allowing two destination nodes to exchange information with each other. Regarding this consideration, increasing attention has been paid to the SWIPT based two-way relay networks (TWRNs), where  wireless signals are  either switched in the time domain or split in the power domain to facilitate SWIPT, i.e., time switching (TS) scheme and power splitting (PS) scheme.

Some studies on the design of TS/PS scheme for two-step TWRNs \cite{T21,T23}  have been hitherto reported.
The authors of \cite{T21} studied the optimal TS/PS scheme for amplify-and-forward (AF) and decode-and-forward (DF) based TWRNs. It was shown that at high signal-to-noise (SNR) ratio, the PS scheme  can achieve a larger sum rate than the TS scheme.
The authors of   \cite{T23} proposed a resource allocation strategy, which jointly optimizes the time allocation ratio and the PS/TS ratio, to minimize the outage probability of DF based TWRNs.

Since  the low complexity of hardware is very vital to  energy-constrained networks,
 three-step two-way relaying has attracted extensive research interests \cite{T31,EL,8287997,2017CL}.
 Based on the TS receiver, three wireless power transfer policies have been proposed to maximize the capacity \cite{T31}.
   A static  equal PS scheme has been developed  to maximize the overall outage capacity for  three-step AF TWRNs \cite{EL}, where the PS ratio is determined by the statistical channel state information (CSI).
The outage capacity can be improved by adopting a   dynamic PS scheme, because the PS
ratio can be adaptive to  the instantaneous CSI instead
of to the  statistical CSI. For this reason, the dynamic equal PS scheme was further developed \cite{8287997}.
Recently, the three-step AF relay has also been  extended  to the DF relay system  \cite{2017CL}, where the upper and lower bounds of the outage probability with respect to the   static equal PS scheme were studied.
Note that although the instantaneous CSI is required at
both destinations to perform successive interference cancellation (SIC) \cite{2017CL}, it is not used in determining the PS ratio.
Moreover, as the channel gains between the source nodes and the relay are both  heterogeneous and instantaneously changing, a PS scheme based on  both heterogeneous and instantaneous CSI can achieve more efficient transmission than the equal PS scheme.

Motivated by the reasons stated above, we propose a dynamic heterogeneous PS scheme, where the PS ratio for each link can be dynamically adjusted based on its instantaneous CSI, and apply it into the SWIPT based three-step DF TWRNs. Unlike the works mentioned above \cite{T31,EL,8287997,2017CL}, we consider a non-linear energy harvesting (EH) model \cite{8355777} instead of the conventional linear one and study the outage capacity of the proposed scheme in the investigated network.

The main contributions of this paper are summarized as follows.
\begin{itemize}
  \item We propose a novel dynamic heterogeneous PS scheme to maximize the capacity of SWIPT based DF TWRNs with the non-linear EH model and  derive the closed-form expressions for the optimal PS ratios. 
      Compared with the scheme in \cite{2017CL}, the proposed scheme is more flexible and can achieve better performance.
  \item  We  analyze the outage capacity with the optimal PS ratios, as an effort  to know how much performance gain the designed scheme could offer in the investigated network.
      Simulation results verify the correctness of the derived results and demonstrate that the proposed PS scheme can significantly improve the capacity of the investigated system as compared with the existing schemes.
  \end{itemize}

The remainder of this paper is organized as follows. The system model is provided in Section \uppercase\expandafter{\romannumeral 2}. In Section \uppercase\expandafter{\romannumeral 3}, we propose a novel dynamic heterogeneous PS scheme to maximize the capacity of SWIPT based DF TWRNs with the non-linear EH model and  analyze the corresponding outage performance.
Simulation results are provided in Section \uppercase\expandafter{\romannumeral 4}, followed by conclusions in Section \uppercase\expandafter{\romannumeral 5}.

\section{System model}
\begin{figure}
  \centering
  \includegraphics[width=0.45\textwidth]{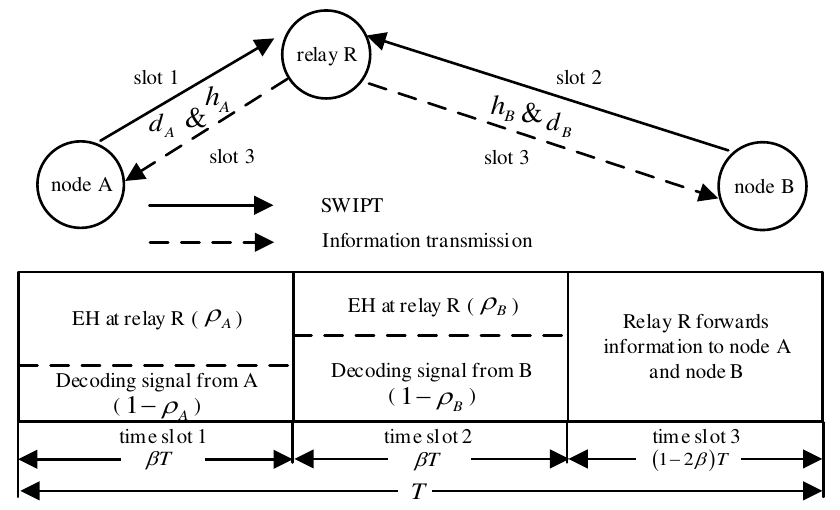}\\
  \caption{System model of the three-step two-way DF relay network.}\label{fig0}
\end{figure}
As shown in Fig. 1, we consider a three-step two-way DF relay network,
where two destination nodes \emph{A} and \emph{B} exchange information via an energy-constrained relay node \emph{R}.
Each node is equipped with a single antenna and works in the half-duplex mode.
There is no direct link between \emph{A} and \emph{B} due to severe path loss and shadowing.
The path loss model is given by  $|h_{i}|^2d_{i}^{ - \alpha }$ ($i = A \;\rm{or}\; B$), where  $h_{i}$ is the $i$-$R$ channel coefficient, ${d_{ i}}$ is the $i$-$R$ distance, and $\alpha $ is the path loss exponent.
All the channels are assumed to undergo independent identically distributed (i.i.d)
quasi-static Rayleigh fading and all the channels are assumed to be reciprocal.
Note that the use of such channels can be found widely in prior research in this field \cite{YeICC,ShiTvt,2017CL}.
Let $T$ denote the total transmission block which can be divided into three time slots.
Let $\beta\in(0,0.5)$ be the time proportion for $R$ to harvest energy and decode signals from $A$ or $B$. The transmission time for $A$ or $B$ to $R$ is $\beta T$. After receiving signal from $i$ ($i = A \;\rm{or}\; B$), $R$ splits it into two parts with ratio $\rho_{i}$ with one part used for energy harvesting and the other part used for information processing.  In the remaining block time $(1-2\beta)T$,  $R$ decodes the signals and forwards them to $A$ and $B$.

At the first or the second time slot with $\beta T$, $A$ or $B$ transmits the signal $s_{A}$ or $s_{B}$ to $R$  and the received RF signal from $i$ ($i = A \;\rm{or}\; B$) at $R$ is given by
\begin{align}\label{1}
{y_{ iR}} = {h_{ i}}\sqrt {{P_{ i}}d_{i}^{-\alpha}} {s_{ i}} + {n_{ iR}},
\end{align}
where $P_{i}$ denotes the transmit power of $i$,
$\mathbb{E}\left\{ {{{\left| s_{i} \right|}^2}} \right\} = 1$ and ${n_{iR}} \sim {\rm{{\cal C}{\cal N}}}\left( {0,\sigma _{iR}^2} \right)$ is the additive white Gaussian noise (AWGN).

Thus, the received power from $i$ at $R$ before the third time slot is given by
\begin{align}\label{2}
{P^{i}_{\rm{RF}}} = {\rho _i}{P_i}|{h_i}{|^2}d_i^{ - \alpha }.
\end{align}
Since the conventional linear EH model cannot capture the practical EH circuit due to the nonlinearity of the diodes, inductors and capacitors \cite{Ng}, we employ a more practical non-linear EH model in \cite{8355777}, which has been verified by comparing with measurement data from \cite{Data1} and \cite{Data2}. Compared with the non-linear EH model in \cite{Ng}, the model in \cite{8355777}, namely piecewise linear EH model, is more mathematically tractable, and able to provide sufficient precision by selecting the proper number of segments (see Fig. 2 in \cite{8355777}).
According to the piecewise linear EH model in \cite{8355777}, the harvested power $P^{i}_{\rm{H}}$ from $i$ can be modelled as
\begin{align}\label{3}
{P^{i}_{\rm{H}}} = \left\{ {\begin{array}{*{20}{c}}
\!\!\!\!\!\!\!\!\!\!\!\!\!\!\!\!\!\!\!\!\!\!{{\rm{0}},\;\;\;\;\;\;\;\;\;\;\;\;\;\;\;{P^{i}_{\rm{RF}}} < P_{\rm{th}}^1};\\
\!\!\!\!{{a_j}{P^{i}_{\rm{RF}}} + {b_j},{P^{i}_{\rm{RF}}} \in \left[ {P_{\rm{th}}^j,P_{\rm{th}}^{j + 1}} \right]}\\
\!\!\!\!\!\!\!\!\!\!\!\!\!\!\!\!\!\!\!\!{{P_{\rm{m}}},\;\;\;\;\;\;\;\;\;\;\;\;{P^{i}_{\rm{RF}}} > {P^N_{\rm{th}}}},
\end{array}} \right.\!\!\!,j = 1,\cdots,N\!-\!\!1;
\end{align}
where ${P_{\rm{th}}} = \{ P_{\rm{th}}^j|1 \le j \le N\} $ are the thresholds on $P^{i}_{\rm{RF}}$ for $N+1$ linear segments\footnote{Note that the $P_{\rm{th}}^1$ represents the power sensitivity for the EH circuits.},
$a_{j}$ and $b_{j}$ are the scope and the intercept for the linear function in the $j$-th segment, respectively, and $P_{\rm{m}}$ denotes the maximum harvestable power when the circuit is saturated.

Then, the total harvested energy is\footnote{If $i=A$, $\bar i=B$; if $i=B$, $\bar i=A$.}
\begin{align}\label{R1}\notag
&E_{\rm{total}}=\beta T (P ^{i}_{\rm{H}}+P^{\overline{i}}_{\rm{H}})\\
&=\beta T\left( {{a_j}{{\rho _i}{P_i}|{h_i}{|^2}d_i^{ - \alpha }} + {a_k} {{\rho _{\bar i}}{P_{\bar i}}|{h_{\bar i}}{|^2}d_{\bar i}^{ - \alpha }} + {b_k} + {b_j}} \right),
\end{align}
where $j,k\in \{0,\cdots,N\}$ denote to which segment ${P^{i}_{\rm{RF}}}$ or ${P^{\overline{i}}_{\rm{RF}}}$ belongs. Let the segment with ${P^{i}_{\rm{RF}}} < P_{\rm{th}}^1$ be the $0$-th segment and the segment with ${P^{i}_{\rm{RF}}} > P_{\rm{th}}^N$ be the $N$-th segment.
Based on Eq. \eqref{3}, we have $a_{0}=b_{0}=a_{N}=0$, $b_{N}=P_{\rm{m}}$. According to \cite{8355777}, both $\{a_{j}\}^{N-1}_{1}$ and $\{b_{j}\}^{N-1}_{1}$ are obtained by linear regression to minimize the difference with the practical EH circuit. Thus, $a_{j}$ and $b_{j}$ in $j$-th $(j = 1,\cdots,N\!-\!\!1)$ segment are given by
\begin{eqnarray}
\begin{cases}
{{a_j} = \frac{{\sum\limits_{l = 1}^n {({x_l} - \bar x)} ({y_l} - \bar y)}}{{\sum\limits_{l = 1}^n {{{({x_l} - \bar x)}^2}} }} = \frac{{\sum\limits_{l = 1}^n {{x_l}} {y_l} - n\bar x\bar y}}{{\sum\limits_{l = 1}^n {x_l^2}  - n{{\bar x}^2}}}};\\
{{b_j} = \bar y - {a_j}\bar x},
\end{cases}
\end{eqnarray}
where $\{(x_{1},y_{1}),(x_{2},y_{2})...,(x_{n-1},y_{n-1}),(x_{n},y_{n})\}$ denote the experimental data in the $j$-th segment, $\overline{x}=\frac{\sum_{l=1}^{n}x_{l}}{n}$, and $\overline{y}=\frac{\sum_{l=1}^{n}y_{l}}{n}$.

For the information processing, 
the received SNR for decoding $s_{i}$ is
\begin{align}\label{5}
{\gamma _{iR}} = \frac{{{P_i}|{h_i}{|^2}\left( {1 - {\rho _i}} \right)}}{{d_i^\alpha \sigma _i^2}}.
\end{align}
Let $\widetilde{s}_{A}$ and $\widetilde{s}_{B}$ denote the decoded signals for $A$ and $B$ during the first and the second time slots, respectively.
In the third time slot, $R$ combines $\widetilde{s}_{A}$ and $\widetilde{s}_{B}$ and broadcasts the normalized signal $s_{R}=\frac{\widetilde{s}_{A}+\widetilde{s}_{B}}{\sqrt{2}}$ to both $A$ and $B$ with the harvested energy $E_{\rm{total}}$. Then the received signal at $i$ is given by
\begin{align}\label{6}\notag
{y_{Ri}} &= {h_i}\sqrt {{P_R}d_i^{ - \alpha }} {s_R} + {n_{Ri}}\\
&\overset{\text{(a)}}{=}{h_i}\sqrt {{P_R}d_i^{ - \alpha }} {\frac{\widetilde{s}_{\overline{i}}}{\sqrt{2}}} + {\widetilde{n}_{Ri}},
\end{align}
where ${P_R} = \frac{{{E_{total}}}}{{\left( {1 - 2\beta } \right)T}}$ is the transmit power at $R$, ${n_{Ri}}=\widetilde{n}_{Ri} \sim {\rm{{\cal C}{\cal N}}}\left( {0,\sigma _{Ri}^2} \right)$ is the AWGN caused by the receiving antenna at $i$,
(a) follows by using SIC due to the fact that the CSI and other system parameters are available at $i$, and $\bar i$ denotes the index of the other destination node.

For analytical simplicity, we assume $P_{A}=P_{B}=P$ and $\sigma _{AR}^2 = \sigma _{BR}^2 = \sigma _{RA}^2 = \sigma _{RB}^2 ={\sigma ^2}$. Based on Eq. \eqref{6}, the end-to-end SNR of the link $\bar i\mathop  \to \limits^R i$ is given by
\begin{align}\label{7}\notag
{\gamma _{Ri}} &= \frac{{{P_R}|{h_i}{|^2}}}{{2d_i^\alpha \sigma ^2}}
 = {a_j}{\rho _i}P|{h_i}{|^4}Xd_i^{ - 2\alpha }  \\
 &+Xd_i^{ - \alpha }\left( { {a_k}{\rho _{\bar i}}P|{h_{\bar i}}{|^2}|{h_i}{|^2}d_{\bar i}^{ - \alpha } + ({b_k} + {b_j})|{h_i}{|^2}} \right),
\end{align}
where ${X} = \frac{\beta }{{2\left( {1 - 2\beta } \right)\sigma _{}^2}}$.

\section{Performance Analysis}
In this section,
we first derive the closed-form expression for the optimal dynamic heterogeneous PS ratios, $\rho^{\ast}_{A}$ and $\rho^{\ast}_{B}$, respectively, to maximize the capacity of the system. Then, an analytical expression of the outage probability with $\rho^{\ast}_{A}$ and $\rho^{\ast}_{B}$ under the piecewise linear EH model is provided.

\subsection{Dynamic Heterogeneous PS Scheme}
Let $\mathbb{P} \left(  \cdot  \right)$ denote the probability. Let $P_{\rm{out}}^{i}$ be the outage probability at  node $i$.
For a predefined threshold ${\gamma _{\rm{th}}}$, $P_{\rm{out}}^{i}$ is given by
\begin{small}
\begin{align}\label{B1}
P_{\rm{out}}^{i}=\underbrace {\mathbb{P}\left( {{\gamma _{\overline{i}R}} < \gamma _{\rm{th}}} \right)}_{P_{1}}+\underbrace {\mathbb{P}\left( {{\gamma _{Ri}} < \gamma _{\rm{th}}},{{\gamma _{\overline{i}R}} \geq\gamma _{\rm{th}}}\right)}_{P_{2}},
\end{align}
\end{small}where $P_{1}$ is the outage probability at the relay and $P_{2}$ is the outage probability at the destination node $i$.

According to \cite{2017CL,Tcom}, the capacity of the system can be calculated as
\begin{small}
\begin{align}\label{B2}
C_{\rm{total}}=\left(2-P_{\rm{out}}^{A}-P_{\rm{out}}^{B}\right)UT\times\min\left(\beta,1-2\beta\right)
\end{align}
\end{small}where $U=\log_{2}(1+{\gamma _{\rm{th}}})$ is the source transmission rate of nodes $A$, $B$, and $R$.
Then we formulate the optimization problem to maximize the capacity of the system as
\begin{small}
\begin{align}\label{B3}
\begin{array}{*{20}{l}}
{P1:\mathop {{\rm{maximize}}}\limits_{({\rho _A},{\rho _B})} \;\;{C_{\rm{total}}}\;}\\
{{\rm{s.t.}}\;:\;\;0 \le {\rho _i} < 1,i \in \left\{ {A,B} \right\}.}
\end{array}
\end{align}
\end{small}

Based on Eq. \eqref{B1},
the optimization problem can be transformed into
\begin{small}
\begin{align}\label{O2}
\begin{array}{*{20}{l}}
{P2:\mathop {{\rm{maximize}}}\limits_{({\rho _A},{\rho _B})} \;\;\mathbb{P}\left( {{\gamma _{RA}} \ge {\gamma _{\rm{th}}}} \right) + \mathbb{P}\;\left( {{\gamma _{RB}} \ge {\gamma _{\rm{th}}}} \right)}\\
{{\rm{s.t.}}\;:\;0 \le {\rho _i} \leq \max\left\{1 - \frac{{{\gamma _{\rm{th}}}d_i^\alpha {\sigma ^2}}}{{P{{\left| {{h_i}} \right|}^2}}},0\right\}, i \in \left\{ {A,B} \right\}}.\!\!\!
\end{array}
\end{align}
\end{small}Note that $P_{1}=1$ always holds when $1 - \frac{{\gamma _{\rm{th}}d_i^\alpha {\sigma ^2}}}{{P{{\left| {{h_i}} \right|}^2}}}<0$.
Since both ${\gamma _{RA}}$ and ${\gamma _{RB}}$ increase with the increasing of ${\rho _i}$ with a given ${\rho _{\overline{i}}}$, it is readily seen that the optimal solution to $P{2}$ can be obtained when $\rho _A = \max \left\{ {1 - \frac{{\gamma _{\rm{th}}d_A^\alpha {\sigma ^2}}}{{P{{\left| {{h_A}} \right|}^2}}},0} \right\}$ and $\rho _B = \max \left\{ {1 - \frac{{\gamma _{\rm{th}}d_B^\alpha {\sigma ^2}}}{{P{{\left| {{h_B}} \right|}^2}}},0} \right\}$.
Thus, the optimal dynamic PS ratio $\rho^{\ast}_{i}$ is given by
\begin{small}
\begin{align}\label{9}
\rho _i^* = \max \left\{ {1 - \frac{{\gamma _{\rm{th}}d_i^\alpha {\sigma ^2}}}{{P{{\left| {{h_i}} \right|}^2}}},0} \right\}.
\end{align}
\end{small}
\subsection{End-to-End Outage Probability with $\rho^{\ast}_{i}$}
Based on Eq. \eqref{B1}, $P_{\rm{out}}^{B}$ can be expressed as
\begin{align}\label{R2}
P_{\rm{out}}^{B}=\underbrace {\mathbb{P}\left( {{\gamma _{AR}} < \gamma _{\rm{th}}} \right)}_{P_{31}}+\underbrace {\mathbb{P}\left( {{\gamma _{RB}} < \gamma _{\rm{th}}},{{\gamma _{AR}} \geq\gamma _{\rm{th}}}\right)}_{P_{32}},
\end{align}
where $P_{31}$ is the outage probability at relay $R$ and $P_{32}$ is the outage probability at destination $B$.
Substituting the optimal PS ratios in Eq. \eqref{9} into Eq. \eqref{R2}, we have
\begin{align}\label{R3}
P_{31}=\mathbb{P}\left({\left| {{h_A}} \right|^2} < \frac{{\gamma _{\rm{th}}d_A^\alpha {\sigma ^2}}}{P} \right)\overset{\text{(b)}}{=}1-\exp\left( { - \frac{\varpi{d_A^\alpha }}{{{\lambda _A}}}} \right),
\end{align}
where (b) holds due to ${\left| {{h_A}} \right|^2} \sim \exp \left( {\frac{1}{{\lambda _A}}} \right)$ and $\varpi= { \frac{{\gamma _{\rm{th}} {\sigma ^2}}}{P}} $.

There are two cases for the value of $\rho _B^*$, which are $0$ for the case with $ 1 - \frac{{\gamma _{\rm{th}}d_B^\alpha {\sigma ^2}}}{{P{{\left| {{h_B}} \right|}^2}}}<0$ and $ 1 - \frac{{\gamma _{\rm{th}}d_B^\alpha {\sigma ^2}}}{{P{{\left| {{h_B}} \right|}^2}}}$ for the other case. Combining with the piecewise linear EH model in Eq. (3), there are
$N + 1$ pairs for values of $(a_{j},b_{j})$.
Thus, $P_{32}$ is given by
\begin{small}
\begin{align}\label{R5}
{P_{32}}
&=\underbrace {\sum\limits_{k = 0}^N  P_{321}^k}_{{P_{21}}} + \underbrace {\sum\limits_{j = 0}^N {\sum\limits_{k = 0}^N {P_{322}^{j,k}} } }_{{P_{22}}},
\end{align}
\end{small}where $P^{k}_{321}$ is the part of $P_{32}$ where the energy
harvester operates in the $k$-th linear region for $P^{A}_{\rm{RF}}$ with $\rho _B^*=0$ and $P^{j,k}_{322}$ is the part of $P_{32}$ where the energy
harvester operates in the $k$-th linear region for $P^{A}_{\rm{RF}}$ and the $j$-th linear region for $P^{B}_{\rm{RF}}$ with $\rho _B^*=1 - \frac{{\gamma _{\rm{th}}d_B^\alpha {\sigma ^2}}}{{P{{\left| {{h_B}} \right|}^2}}}$.

Based on the above two cases, if ${\left| {{h_B}} \right|^2} < \varpi d_B^\alpha $ is satisfied, we have $P^{k}_{321}$ as
\begin{small}
\begin{align}\label{G1}\notag
&P^{k}_{321}=\\
&\mathbb{P}\!\!\left( {{a_k}|{h_A}{|^2} < \frac{{Y^{A1} }}{{{{\left| {{h_B}} \right|}^2}}} + Y_{0,k}^{A3},{{\left| {{h_A}} \right|}^2} \ge \varpi d_A^\alpha ,{{\left| {{h_B}} \right|}^2} < \varpi d_B^\alpha } \right),\!\!\!\!\!
\end{align}
\end{small}where 
${Y^{A1}} = \frac{{{\gamma _{th}}d_B^\alpha d_A^\alpha }}{{PX}}$ and ${Y^{A3}_{j,k}} =[\varpi \left( {{a_k} + {a_j}} \right)-\frac{{b_k} + {b_j}}{P}]d_A^\alpha $.
If ${\left| {{h_B}} \right|^2} \geq \varpi d_B^\alpha $, $P^{j,k}_{322}$ are
\begin{small}
\begin{align}\label{G2}\notag
&P^{j,k}_{322}=\\
&\mathbb{P}\!\!\left(\!\!{a_k}\varpi d_A^\alpha\leq{a_k}|{h_A}{|^2} \!<\!\! \frac{{{Y^{A1}}}}{{|{h_B}{|^2}}}\! +\! {Y^{A2}_{j}}|{h_B}{|^2} \!+ \!{Y^{A3}_{j,k}} ,{\left| {{h_B}} \right|^2}\!\! \ge\! \varpi d_B^\alpha \!\!\right),
\end{align}
\end{small}where ${Y^{A2}_{j}} =  - \frac{{{a_j}d_A^\alpha }}{{d_B^\alpha }}$.

\subsubsection{The derivation of $P_{21}$}
Based on the value of $k$ and the piecewise linear EH model in Eq. (3), there are three cases for $P^{k}_{321}$ as follows.

\textbf{Case I:}
When $k=0$, we have $a_{k}=b_{k}=0$ and $P^{A}_{\rm{RF}}<P^{1}_{\rm{th}}$.
Combining with ${\left| {{h_B}} \right|^2} < \varpi d_B^\alpha $ is given by
\begin{small}
\begin{align}\label{R10}\notag
&P^{0}_{321}=\mathbb{P}\!\left(\!0 <{Y^{A1}}, \varpi d_A^\alpha\!\leq\!{\left| {{h_A}} \right|^2}\! < \!(\varpi\!+\!\theta_{1}) d_A^\alpha,{\left| {{h_B}} \right|^2}\! <\! \varpi d_B^\alpha \right)\\
&\overset{\text{(c)}}{=}\!\left[ {1 - \!\exp\!\! \left( { - \frac{{\varpi d_B^\alpha }}{{{\lambda _B}}}} \right)} \!\right]\left[ {\exp \!\left(\! { - \frac{{\varpi d_A^\alpha }}{{{\lambda _A}}}}\! \right) - \exp  \left(\! { - \frac{{({\theta _1} \!+ \varpi )d_A^\alpha }}{{{\lambda _A}}}} \!\right)} \right],\!\!
\end{align}
\end{small}where (c) holds due to ${\left| {{h_B}} \right|^2} \sim \exp \left( {\frac{1}{{\lambda _B}}} \right)$ and $\theta_{1}=\frac{P^{1}_{\rm{th}}}{P}$.

\textbf{Case II:}
When $k=N$, we have $a_{k}=0, b_{k}=P_{\rm{m}}$. Let $\theta_{k}=\frac{P^{k}_{\rm{th}}}{P}$. Then ${\left| {{h_A}} \right|^2}> (\varpi+\theta_{N}){d_A^\alpha }$ and $P^{N}_{321}$ is given by
\begin{small}
\begin{align}\label{R12}\notag
P^{N}_{321}&=\mathbb{P}\left(0\! <\!\frac{{Y^{A1}}}{{{{\left| {{h_B}} \right|}^2}}}+ \!{Y^{A3}_{0,N}} , {\left| {{h_A}} \right|^2} \!> \!(\varpi+\theta_{N}) d_A^\alpha, {\left| {{h_B}} \right|^2} <\!\! \varpi d_B^\alpha \!\!\right)\\
&=\exp \left( { - \frac{{(\varpi  + {\theta _N})d_A^\alpha }}{{{\lambda _A}}}} \right)\left[\!1-\!\exp\left(\!\! { - \frac{\delta^{0}}{{{\lambda _B}}}} \right)\right],
\end{align}
\end{small}where $\delta^{0}=\min\left(\frac{{{\gamma _{\rm{th}}}d_B^\alpha }}{{{P_{\rm{m}}}X}},{\varpi d_B^\alpha }\right)$.

\textbf{Case III:}
When $k \in \{ 1,...N - 1\} $, we have $a_{k}\neq 0$ and $(\varpi+\theta_{k}){d_A^\alpha }\leq{\left| {{h_A}} \right|^2} \leq (\varpi+\theta_{k+1})d_A^\alpha$.
Based on Eq. \eqref{G1}, $P^{k}_{321}$ is given by
\begin{small}
\begin{align}\label{R11}\notag
&P^{k}_{321}=\mathbb{P}\left({\left| {{h_B}} \right|^2} < \varpi d_B^\alpha , (\varpi+\theta_{k}){d_A^\alpha }\leq{\left| {{h_A}} \right|^2} \leq {\phi^{k} _{\max }}(\left| {{h_B}} \right|^2) \right)\\ \notag
&\overset{\text{(d)}}{=}\int_0^{\varpi d_B^\alpha } {\int_{(\varpi  + {\theta _k})d_A^\alpha }^{\phi _{\max }^k\left( x \right)} {\frac{1}{{{\lambda _B}}}\exp \left( { - \frac{x}{{{\lambda _B}}}} \right)} } \frac{1}{{{\lambda _A}}}\exp \left( { - \frac{y}{{{\lambda _A}}}} \right)dydx\\ \notag
&=\exp \left[ { - \frac{{(\varpi  + {\theta _k})d_A^\alpha }}{{{\lambda _A}}}} \right] - \exp \!\left[ { - \frac{{(\varpi  + {\theta _k})d_A^\alpha }}{{{\lambda _A}}}} \!\right]\exp \!\left( \!{ - \frac{{\delta _{\max }^k}}{{{\lambda _B}}}} \!\right)\\ \notag
&\;\;\;\;-\exp \left[ { - \frac{{(\varpi  + {\theta _{k + 1}})d_A^\alpha }}{{{\lambda _A}}}} \right]\left[ {1 - \exp \!\left( { - \frac{{\delta _{\min }^k}}{{{\lambda _B}}}} \right)} \right]\\
&\;\;\;\;-\frac{{\exp \left( { - \frac{{\varpi d_A^\alpha }}{{{\lambda _A}}}} \right)}}{{{\lambda _B}}}\underbrace {\int_{\delta _{\min }^k}^{\delta _{\max }^k} {\exp \left( { - \frac{Y^{A1}}{{{a_k}{\lambda _A}x}} - \frac{x}{{{\lambda _B}}}} \right)}}_{\Xi},
\end{align}
\end{small}where (d) holds from $y={{{\left| {{h_A}} \right|}^2}}$ and $x={{{\left| {{h_B}} \right|}^2}}$ and
\begin{small}
\begin{align}\notag
\left\{ {\begin{array}{*{20}{c}}
{\!\!\!\!\!\!\!\!\!\!\!\!\!\!\!\!\!\!\!\!\!\!\!\!\!\!\!\!\!\!\!\!\!\!\!\!\!\!\!\!\!\!\!\!\!\!\!\!\!\!\!\!\!\!\!\!\!\!\!\!\!\!\!\!\!\!\!\!\!\!\!\!\!\!\!\!\!\!\!\!\!\!\!\!\!\!\!\!\!\!\!\!\!\!\!\!\!\!\!\!\!\!\!\!\!\!\!\!\!\!\!\!\!\!\!\phi _{\max }^k(x){\rm{ = }}}\\
{\max \left[ {\min \left( {(\varpi  + {\theta _{k + 1}})d_A^\alpha ,\frac{{{Y^{A1}}}}{{{a_k}x}} + \varpi d_A^\alpha } \right),(\varpi  + {\theta _k})d_A^\alpha } \right]},\\
{\!\!\!\!\!\!\!\!\!\!\!\!\!\!\!\!\!\!\!\!\!\!\!\!\!\!\!\!\!\!\!\!\!\!\!\!\!{\delta ^k_{{{\min }}}} = \min \left[ {\max \left( {\frac{{{Y^{A1}}}}{{d_A^\alpha {a_k}{\theta _{k + 1}}}},0} \right),\varpi d_B^\alpha } \right],}\\
{\!\!\!\!\!\!\!\!\!\!\!\!\!\!\!\!\!\!\!\!\!\!\!\!\!\!\!\!\!\!\!\!\!{\delta ^k_{{{\max }}}} = \max \left[ {\min \left( {\frac{{{Y^{A1}}}}{{d_A^\alpha {a_k}{\theta _k}}},\varpi d_B^\alpha } \right),{\delta^k _{{{\min }}}}} \right].}
\end{array}} \right.
\end{align}
\end{small}

Note that it is difficult to find the accurate closed-form expression for $P^{k}_{321}$ with ${Y_{0,k}^{A1}}>0$ due to the integral $\int_{{s_1}}^{{s_2}} {\exp ({z_1}x + \frac{{{z_2}}}{x})} dx$ with any value of $z_{1}$ and $z_{2}\neq 0$. Fortunately, we can use Gaussian-Chebyshev quadrature to  find an approximation for $P^{k}_{321}$.
According to \cite{YeICC}, Gaussian-Chebyshev quadrature is defined as $\int_{t_{1}}^{t_{2}} {f\left( \xi \right)d\xi \approx \frac{{t_{2} - t_{1}}}{2}\sum\limits_{j = 1}^K {{w_j}\sqrt {1 - z_j^2} f\left( {{\xi_j}} \right)} } $, where ${w_j} = \frac{\pi }{K}$, ${z_j} = \cos \frac{{2j - 1}}{{2K}}\pi $ and ${\xi_j} = \frac{{t_{2}- t_{1}}}{2}{z_j} + \frac{{t_{2} + t_{1}}}{2}$. Thus, $\Xi$ can be calculated as
\begin{small}
\begin{align}\label{19-1}
&\Xi \approx \frac{{\pi (\delta _{\max }^k - \delta _{\min }^k)}}{{2M}}\!\!\sum\limits_{m = 1}^M\!\! {\sqrt {1 - \nu _m^2} } \exp \!\! \left(\!\! { - \frac{Y^{A1}}{{{a_k}{\lambda _A}\kappa _m^k}} - \frac{{\kappa _m^k}}{{{\lambda _B}}}} \!\!\right),\!\!
\end{align}
\end{small}where $M$ is a parameter that determines the tradeoff between complexity and accuracy, ${\nu _m} = \cos \frac{{2m - 1}}{{2M}}\pi $, and $\kappa _m^{k} = \frac{(\delta _{\max }^k - \delta _{\min }^k)}{2}{\nu _m} + \frac{(\delta _{\max }^k + \delta _{\min }^k)}{2}$.
Note that a larger $M$ results in a higher accuracy while a moderate yet acceptable accuracy can be realized at a small $M$. This  is verified in our simulation results.
Based on Eqs. (\ref{R10}), (\ref{R12}), and (\ref{R11}), the approximation of $P_{21}$ can be obtained.

\subsubsection{The derivation of $P_{22}$}
Likewise, we derive the expression of $P_{22}$ as follows.
Given the values of $j$ and $k$, $P^{j,k}_{322}$ is expressed as follows.

\textbf{Case 1:}
When $j=0$ with $\rho _B^*=1 - \frac{{\gamma _{\rm{th}}d_B^\alpha {\sigma ^2}}}{{P{{\left| {{h_B}} \right|}^2}}}$, we have $a_{j}=b_{j}=0$ and $\varpi d_B^\alpha \leq{\left| {{h_B}} \right|^2}< (\theta_{1}+\varpi) d_B^\alpha$. Based on Eq. \eqref{G2}, the expression of $P_{322}^{0,k}$ is given by
\begin{small}
\begin{align}\label{G3}\notag
&P^{0,k}_{322}=\\
&\!\!\mathbb{P}\!\!\left(\!{a_k}|{h_A}{|^2} < \frac{Y^{A1}}{{{{\left| {{h_B}} \right|}^2}}} + Y_{0,k}^{A3}\!\!,{\left|\! {{h_A}}\! \right|^2} \!\!\ge\!\! \varpi d_A^\alpha\!\!, \varpi d_B^\alpha \!\leq\!{\!\left|\! {{h_B}}\! \right|^2}\!\!<\!\! (\theta_{1}\!\!+\!\varpi) d_B^\alpha \!\!\right).\!\!\!
\end{align}
\end{small}Similarly, when $k=0$, we have
\begin{small}
\begin{align}\notag
P_{322}^{0,0}&=\left[ { \exp \left( { - \frac{{\varpi d_B^\alpha }}{{{\lambda _B}}}}\right)}-\exp \left( { - \frac{{(\theta_{1}+\varpi) d_B^\alpha }}{{{\lambda _B}}}} \right) \right]\\ \notag
&\;\;\;\;\times \left[ {\exp \!\left( { - \frac{{\varpi d_A^\alpha }}{{{\lambda _A}}}}\! \right) - \exp  \left( { - \frac{{({\theta _1} + \varpi )d_A^\alpha }}{{{\lambda _A}}}} \!\right)} \right].
\end{align}
\end{small}
 When $k \in \{ 1,...N - 1\} $, $P_{322}^{0,k}$ can be calculated as
\begin{small}
\begin{align}\label{G4} \notag
P_{322}^{0,k}
&\approx\!\exp\!\! \left(\! {\! - \frac{{(\varpi \! + \!{\theta _k})d_A^\alpha }}{{{\lambda _A}}}} { \!- \frac{{\varpi d_B^\alpha }}{{{\lambda _B}}}} \!\!\right)\!-\! \exp \!\left(\! { - \frac{{(\varpi\!  + \! {\theta _k})d_A^\alpha }}{{{\lambda _A}}}} { - \frac{{\delta _{\max }^{0,k}}}{{{\lambda _B}}}}\!\right)\\ \notag
&\;\;\;\;-\!\exp \!\left(\!\! { - \frac{{(\varpi \! +\! {\theta _{k + 1}})d_A^\alpha }}{{{\lambda _A}}}} \!\!\right)\left[ {\exp\! \left(\! { - \frac{{\varpi d_B^\alpha }}{{{\lambda _B}}}} \!\right) \!-\! \exp\! \left(\! { - \frac{{\delta _{\min }^{0,k}}}{{{\lambda _B}}}} \!\right)} \right]\\ \notag
&\;\;\;\;-\!\exp\! \left(\!\! { - \frac{{\varpi d_A^\alpha }}{{{\lambda _A}}}} \!\!\right)\!
\frac{{\pi (\delta _{\max }^{0,k} \!-\! \delta _{\min }^{0,k})}}{{2M\lambda _B}}\!\!\\
&\;\;\;\;\times\sum\limits_{m = 1}^M {\sqrt {1 - \nu _m^2} } \exp  \left( { - \frac{Y^{A1}}{{{a_k}{\lambda _A}\kappa _m^{0,k}}} - \frac{{\kappa _m^{0,k}}}{{{\lambda _B}}}} \right),
\end{align}
\end{small}where
\begin{small}
\begin{align}\notag
\left\{ {\begin{array}{*{20}{c}}
{\delta _{\min }^{0,k} = \min \left[ {\max \left( {\frac{{{Y^{A1}}}}{{d_A^\alpha {a_k}{\theta _{k + 1}}}},\varpi d_B^\alpha } \right),({\theta _1} + \varpi )d_B^\alpha } \right],{\rm{ }}}\\
{\!\!\!\!\!\!\!\!\delta _{\max }^{0,k} = \max \left[ {\min \left( {\frac{{{Y^{A1}}}}{{d_A^\alpha {a_k}{\theta _k}}},({\theta _1} + \varpi )d_B^\alpha } \right),\delta _{\min }^{0,k}} \right],}\\
{\!\!\!\!\!\!\!\!\!\!\!\!\!\!\!\!\!\!\!\!\!\!\!\!\!\!\!\!\!\!\!\!\!\!\!\!\!\!\!\kappa _m^{0,k} = \frac{{(\delta _{\max }^{0,k} - \delta _{\min }^{0,k})}}{2}{\nu _m} + \frac{{(\delta _{\max }^{0,k} + \delta _{\min }^{0,k})}}{2}}.
\end{array}} \right.
\end{align}
\end{small}
When $k=N$, $P_{322}^{0,N}$ is given by
\begin{small}
\begin{align}\label{G5}
&P_{322}^{0,N}=\exp \left( { - \frac{{(\varpi  + {\theta _N})d_A^\alpha }}{{{\lambda _A}}}} \right)\left[\exp\left( \!{ - \frac{\varpi{d_B^\alpha }}{{{\lambda _B}}}} \right)-\!\exp\left( { - \frac{\delta^{0,N}}{{{\lambda _B}}}} \!\!\right)\right],
\end{align}
\end{small}where $\delta^{0,N}=\max \left( {\min \left[ {\frac{{{\gamma _{\rm{th}}}d_B^\alpha }}{{{P_{\rm{m}}}X}},({\theta _1} + \varpi )d_B^\alpha } \right],\varpi d_B^\alpha } \right)$.

\textbf{Case 2:}
When $j=N$, we have $a_{j}=0$ and $b_{j}=P_{m}$. Thus, $P^{N,k}_{322}$ is given by \small{$\mathbb{P}\!\!\left({a_k}|{h_A}{|^2} < \frac{Y^{A1}}{{{{\left| {{h_B}} \right|}^2}}} + Y_{N,k}^{A3}\!\!,{\!\left| {{h_A}} \right|^2} \!\!\ge\! \varpi d_A^\alpha, {\left| {{h_B}} \right|^2} \!\!>\! (\varpi\!+\!\theta_{N})d_B^\alpha  \!\!\right).$}
\normalsize{Similar} to the derivation of $P_{21}$,
if $k=0$, $P^{N,0}_{322}$ can be calculated as
\begin{small}
\begin{align}\label{R18}\notag
P^{N,0}_{322}&=\left[\exp \left( { - \frac{{(\varpi  +\! {\theta _N})d_B^\alpha }}{{{\lambda _B}}}} \right) \!- \exp\left( { - \frac{\delta^{N,0}}{{{\lambda _B}}}} \right)\right]\\
&\;\;\;\;\times\left[\exp\left( { - \frac{\varpi{d_A^\alpha }}{{{\lambda _A}}}} \right)-\exp\left( { - \frac{(\theta_{1}+\varpi){d_A^\alpha }}{{{\lambda _A}}}} \right)\right],
\end{align}
\end{small}where $\delta^{N,0}=\max \left[ {\frac{{{\gamma _{\rm{th}}}d_B^\alpha }}{{{P_{\rm{m}}}X}},({\theta _N} + \varpi )d_B^\alpha } \right]$.

If $k\in \{1,\cdots,N-1\}$, $P^{N,k}_{322}$ can be computed as
\begin{small}
\begin{align}\label{R19}\notag
&P^{N,k}_{322}\\ \notag
&\approx\!\exp\! \left(\! { - \frac{{(\varpi \! + \!{\theta _k})d_A^\alpha }}{{{\lambda _A}}}} { \!- \! \frac{(\varpi\!+\!\theta_{N})d_B^\alpha }{{{\lambda _B}}}} \!\right)\!-\! \exp \!\left( \!\!{ - \frac{{(\varpi \! +\! {\theta _k})d_A^\alpha }}{{{\lambda _A}}}} { \!- \!\frac{{\delta _{\max }^{N,k}}}{{{\lambda _B}}}}\!\!\right)\\ \notag
&\;\;\;\;-\left[ {\exp \left( { - \frac{(\varpi\!+\!\theta_{N})d_B^\alpha }{{{\lambda _B}}}} \right) - \exp \left( { - \frac{{\delta _{\min }^{N,k}}}{{{\lambda _B}}}} \right)} \right]\times\\ \notag
&\;\;\;\;\exp \left( { - \frac{{(\varpi  + {\theta _{k + 1}})d_A^\alpha }}{{{\lambda _A}}}} \right)-\exp \left( { - \frac{{\varpi d_A^\alpha }}{{{\lambda _A}}}} \right)
\frac{{\pi (\delta _{\max }^{N,k} - \delta _{\min }^{N,k})}}{{2M\lambda _B}}\!\!\\
&\;\;\;\;\times\sum\limits_{m = 1}^M\!\! {\sqrt {1 - \nu _m^2} } \exp \!\! \left(\!\! { - \frac{Y^{A1}}{{{a_k}{\lambda _A}\kappa _m^{N,k}}} - \frac{{\kappa _m^{N,k}}}{{{\lambda _B}}}} \right),
\end{align}
\end{small}where
\begin{small}
\begin{align}\notag
\left\{ {\begin{array}{*{20}{c}}
{\delta _{\min }^{N,k}{\rm{ }} = \max \left( {\frac{{{Y^{A1}}}}{{d_A^\alpha {a_k}{\theta _{k + 1}}}},(\varpi  + {\theta _N})d_B^\alpha } \right),{\rm{ }}}\\
{\!\!\!\!\!\!\delta _{\max }^{N,k} = \max \left( {\frac{{{Y^{A1}}}}{{d_A^\alpha {a_k}{\theta _k}}},(\varpi  + {\theta _N})d_B^\alpha } \right),}\\
{\!\!\!\!\kappa _m^{N,k} = \frac{{(\delta _{\max }^{N,k} - \delta _{\min }^{N,k})}}{2}{\nu _m} + \frac{{(\delta _{\max }^{N,k} + \delta _{\min }^{N,k})}}{2}.}
\end{array}} \right.
\end{align}
\end{small}

If $k=N$, $P^{N,N}_{322}$ is given by
\begin{small}
\begin{align}\label{R20}\notag
P^{N,N}_{322}&=\exp \left( { - \frac{{(\varpi  + {\theta _N})d_A^\alpha }}{{{\lambda _A}}}} \!\!\right)\times \\
&\;\;\;\;\left[\exp \left( { - \frac{{(\varpi  + {\theta _N})d_B^\alpha }}{{{\lambda _B}}}} \right) - \exp \left({ - \frac{\delta^{N,N}}{{{\lambda _B}}}} \right)\right],
\end{align}
\end{small} where $\delta^{N,N}=\max \left( {\frac{{{\gamma _{\rm{th}}}d_B^\alpha }}{2{{P_{\rm{m}}}X}},({\theta _N} + \varpi )d_B^\alpha } \right)$.

\textbf{Case 3:}
When the energy harvester for ${P^{B}_{\rm{RF}}}$ works in the $j$-th linear region with $j\in \{1,\cdots,N-1\}$, we have $(\varpi+\theta_{j}){d_B^\alpha }\leq{\left| {{h_B}} \right|^2} \leq (\varpi+\theta_{j+1})d_B^\alpha $.

(1) For the case ${P^{A}_{\rm{RF}}}<P^{1}_{\rm{th}}$, we have $k=0$.
Based on Eq. \eqref{G2}, there are two cases for $P^{j,0}_{322}$.

If $\Delta_{j,0}  = ({Y^{A3}_{j,0}})^{2} - 4{Y^{A1}}Y_j^{A2} < 0$, there is no ${\left| {{h_B}} \right|^2}$ that satisfies $Y_j^{A2}|{h_B}{|^4} + Y_{j,0}^{A3}|{h_B}{|^2} + {Y^{A1}} > 0$ and $P^{j,0}_{322}=0$ always holds.

If $\Delta_{j,0} \geq 0$, there are two solutions to the equation $Y_j^{A2}|{h_B}{|^4} + Y_{j,0}^{A3}|{h_B}{|^2} + {Y^{A1}} = 0$, which are
\begin{small}
\begin{align}\notag
\left\{ {\begin{array}{*{20}{c}}
{x_{j,0}^{B1} = \min \left( {\frac{{ - Y_{j,0}^{A3} - \sqrt {{\Delta _{j,0}}} }}{{2Y_j^{A2}}},\frac{{ - Y_{j,0}^{A3} + \sqrt {{\Delta _{j,0}}} }}{{2Y_j^{A2}}}} \right),}\\
{x_{j,0}^{B2} = \max \left( {\frac{{ - Y_{j,0}^{A3} - \sqrt {{\Delta _{j,0}}} }}{{2Y_j^{A2}}},\frac{{ - Y_{j,0}^{A3} + \sqrt {{\Delta _{j,0}}} }}{{2Y_j^{A2}}}} \right).}
\end{array}} \right.
 \end{align}
 \end{small}Combining with the condition that $(\varpi+\theta_{j}){d_B^\alpha }\leq{\left| {{h_B}} \right|^2} \leq (\varpi+\theta_{j+1})d_B^\alpha $, $P^{j,0}_{322}$ is given by
\begin{small}
\begin{align}\label{R13}\notag
P^{j,0}_{322}&=\mathbb{P}\left(\varpi d_A^\alpha\leq{\left| {{h_A}} \right|^2} < (\theta_{1}+\varpi) d_A^\alpha , {\delta^{j,0} _{\min }}\leq{\left| {{h_B}} \right|^2} \leq {\delta^{j,0} _{\max }} \right)\\ \notag
&=\left(\!\exp\left( { - \frac{\varpi{d_A^\alpha }}{{{\lambda _A}}}} \right)-\exp\left( { - \frac{(\theta_{1}+\varpi){d_A^\alpha }}{{{\lambda _A}}}} \right)\right)\\
&\;\;\;\;\times\left(\exp \left( { - \frac{{{\delta^{j,0} _{\min }}}}{{{\lambda _B}}}}  \right) - \!\exp \!\left( { - \frac{{{\delta^{j,0} _{\max }}}}{{{\lambda _B}}}} \right)\right),
\end{align}
\end{small}where
\begin{align}\notag
\left\{ {\begin{array}{*{20}{c}}
{\delta _{\min }^{j,0} = \min \left( {\max \left( {(\varpi  + {\theta _j})d_B^\alpha ,x_{j,0}^{B1}} \right),(\varpi  + {\theta _{j + 1}})d_B^\alpha } \right),}\\
{\!\!\!\!\!\!\!\!\!\!\!\!\!\!\!\!\delta _{\max }^{j,0} = \max \left( {\min \left( {(\varpi  + {\theta _{j + 1}})d_B^\alpha ,x_{j,0}^{B2}} \right),{\delta ^{j,0}_{{{\min }}}}} \right).}
\end{array}} \right.
\end{align}

(2) For the case $P_{\rm{RF}}^A\in {\rm{ [}}{P_{\rm{th}}^j}, {P_{\rm{th}}^{j + 1}}]$ with $j\in \{1,\cdots,N-1\}$, $P_{322}^{j,k}$ with $j,k \in \{1,\cdots,N-1\}$ is given by
\begin{small}
\begin{align}\label{R14}\notag
\!\!&\!P_{322}^{j,k}\!=\!\\ \notag
&\mathbb{P}\!\left(\! (\varpi \! +\! {\theta _k})d_A^\alpha \!\!\leq{|{h_A}{|^2} \!<\! \phi _{\max }^{j,k}, (\theta_{j}\!+\!\varpi) d_B^\alpha \!\! \le \!{{\left| {{h_B}} \right|}^2} \!\!<\! ({\theta _{j+1}}\! + \!\!\varpi )d_B^\alpha} \!\!\right)\\ \notag
&=\!\left(\! {\exp \!\left(\! { - \frac{{(\varpi  + {\theta _j})d_B^\alpha }}{{{\lambda _B}}}} \!\right)\! -\! \exp \!\left(\! { - \frac{{(\varpi  + {\theta _{j + 1}})d_B^\alpha }}{{{\lambda _B}}}} \!\right)} \!\right)\times \\
&\exp \!\!\left( \!\!{ - \frac{{(\varpi \! + {\theta _k})d_A^\alpha }}{{{\lambda _A}}}} \!\right)\!\!-\! \frac{1}{{{\lambda _B}}}\int_{(\varpi  \!+ {\theta _j})d_B^\alpha }^{(\varpi  \!+\! {\theta _{j + 1}})d_B^\alpha } \!\!\!{\exp \!\left(\! { - \!\frac{{\phi _{\max }^{j,k}\!\left(\! x \!\right)}}{{{\lambda _A}}} - \frac{x}{{{\lambda _B}}}} \right)} dx,
\end{align}
\end{small}where
\begin{small}
\begin{align}\notag
&\phi _{\max }^{j,k}\left( x \right) = \\ \notag
&\max \!\!\left(\!\! {\min \!\left(\! {\frac{{Y^{A1} }}{{{a_k}x}}\! + \!\frac{{Y_j^{A2}}}{{{a_k}}}x \!+ \! \frac{{Y_{j,k}^{A3}}}{{{a_k}}},(\varpi \! +\! {\theta _{k + 1}})d_A^\alpha } \!\!\right),
(\varpi  + {\theta _k})d_A^\alpha } \!\!\right).
\end{align}
\end{small}

By using Gaussian-Chebyshev quadrature, the approximation of $P_{322}^{j,k}$ is given by
\begin{small}
\begin{align}\label{RR}\notag
P_{322}^{j,k}&\approx \!\left(\! {\exp \!\left(\! { - \frac{{(\varpi  + {\theta _j})d_B^\alpha }}{{{\lambda _B}}}} \!\right)\! -\! \exp \!\left(\! { - \frac{{(\varpi  + {\theta _{j + 1}})d_B^\alpha }}{{{\lambda _B}}}} \!\right)} \!\right) \\ \notag
&\;\;\;\;\times\exp \!\!\left( \!\!{ - \frac{{(\varpi \! + {\theta _k})d_A^\alpha }}{{{\lambda _A}}}} \!\right)\!\!-\!\frac{{\pi (\theta_{j+1} - \theta_{j})d^{\alpha}_{B}}}{{2M\lambda_{B}}}\\
&\;\;\;\;\times\sum\limits_{m = 1}^M\!\! {\sqrt {1 - \nu _m^2} } \exp \!\! \left(\!\! { - \frac{\phi _{\max }^{j,k}\left( \kappa _m^{j,k} \right)}{{{\lambda _A}}} - \frac{{\kappa _m^{j,k}}}{{{\lambda _B}}}} \!\!\right),
\end{align}
\end{small}where $\kappa _m^{jk} = \frac{{({\theta _{j + 1}} - {\theta _j})d_B^\alpha }}{2}{\nu _m} + \frac{{(2\varpi  + {\theta _j} + {\theta _{j + 1}})d_B^\alpha }}{2}$.

\normalsize{(3)} For the case ${P^{A}_{\rm{RF}}}>P^{N}_{\rm{th}}$, we have $k=N$ and $a_{k}=0, b_{k}=P_{\rm{m}}$. Based on Eq. \eqref{G2}, the value of $P_{322}^{j,N}$ depends on $\Delta_{j,N}  = ({Y^{A3}_{j,N}})^{2} - 4{Y^{A1}}Y_j^{A2}$. If $\Delta_{j,N}<0$, similar to $P_{322}^{j,0}$, we have $P_{322}^{j,N}=0$. If $\Delta_{j,N}\geq0$, $P_{322}^{j,N}$ is given by
\begin{small}
\begin{align}\label{R15}\notag
P^{j,N}_{322}&=\mathbb{P}\left({\left| {{h_A}} \right|^2} > (\theta_{N}+\varpi) d_A^\alpha , {\delta^{j,N} _{\min }}\leq{\left| {{h_B}} \right|^2} \leq {\delta^{j,N} _{\max }} \right)\\
&=\!\exp\!\left(\!\! { - \frac{(\theta_{N}+\varpi){d_A^\alpha }}{{{\lambda _A}}}} \!\!\right)\!\!\left(\!\exp\! \left( \!\!{ - \frac{{{\delta^{j,N} _{\min }}}}{{{\lambda _B}}}}  \!\right)\!\! - \!\exp \!\left(\! { - \frac{{{\delta^{j,N} _{\max }}}}{{{\lambda _B}}}} \!\!\right)\!\!\right),
\end{align}
\end{small}where
\begin{small}
\begin{align}\notag
\left\{ {\begin{array}{*{20}{c}}
{\delta _{\min }^{j,N} = \min \left( {\max \left( {(\varpi  + {\theta _j})d_B^\alpha ,x_{j,N}^{B1}} \right),(\varpi  + {\theta _{j + 1}})d_B^\alpha } \right),}\\
{\!\!\!\!\!\!\!\!\!\!\!\!\!\!\!\!\delta _{\max }^{j,N} = \max \left( {\min \left( {(\varpi  + {\theta _{j + 1}})d_B^\alpha ,x_{j,N}^{B2}} \right),\delta _{\min }^{j,N}} \right),}\\
{\!\!\!\!\!\!\!\!\!\!\!\!\!\!\!\!\!\!\!\!\!\!\!x_{j,N}^{B1} = \min \left( {\frac{{ - Y_{j,N}^{A3} - \sqrt {{\Delta _{j,N}}} }}{{2Y_j^{A2}}},\frac{{ - Y_{j,N}^{A3} + \sqrt {{\Delta _{j,N}}} }}{{2Y_j^{A2}}}} \right),}\\
{\!\!\!\!\!\!\!\!\!\!\!\!\!\!\!\!\!\!\!\!\!\!x_{j,N}^{B2} = \max \left( {\frac{{ - Y_{j,N}^{A3} - \sqrt {{\Delta _{j,N}}} }}{{2Y_j^{A2}}},\frac{{ - Y_{j,N}^{A3} + \sqrt {{\Delta _{j,N}}} }}{{2Y_j^{A2}}}} \right).}
\end{array}} \right.
\end{align}
\end{small}\normalsize{Based} on Eq. \eqref{R5}, the approximation of $P^{B}_{\rm{out}}$ can be obtained.
Similarly, $P_{\rm{out}}^{A}$ can be obtained by the same way. Based on $P_{\rm{out}}^{A}$ and $P_{\rm{out}}^{B}$, the capacity of the system $C_{\rm{total}}$ can be determined.
\begin{figure}
  \centering
  \includegraphics[width=0.47\textwidth]{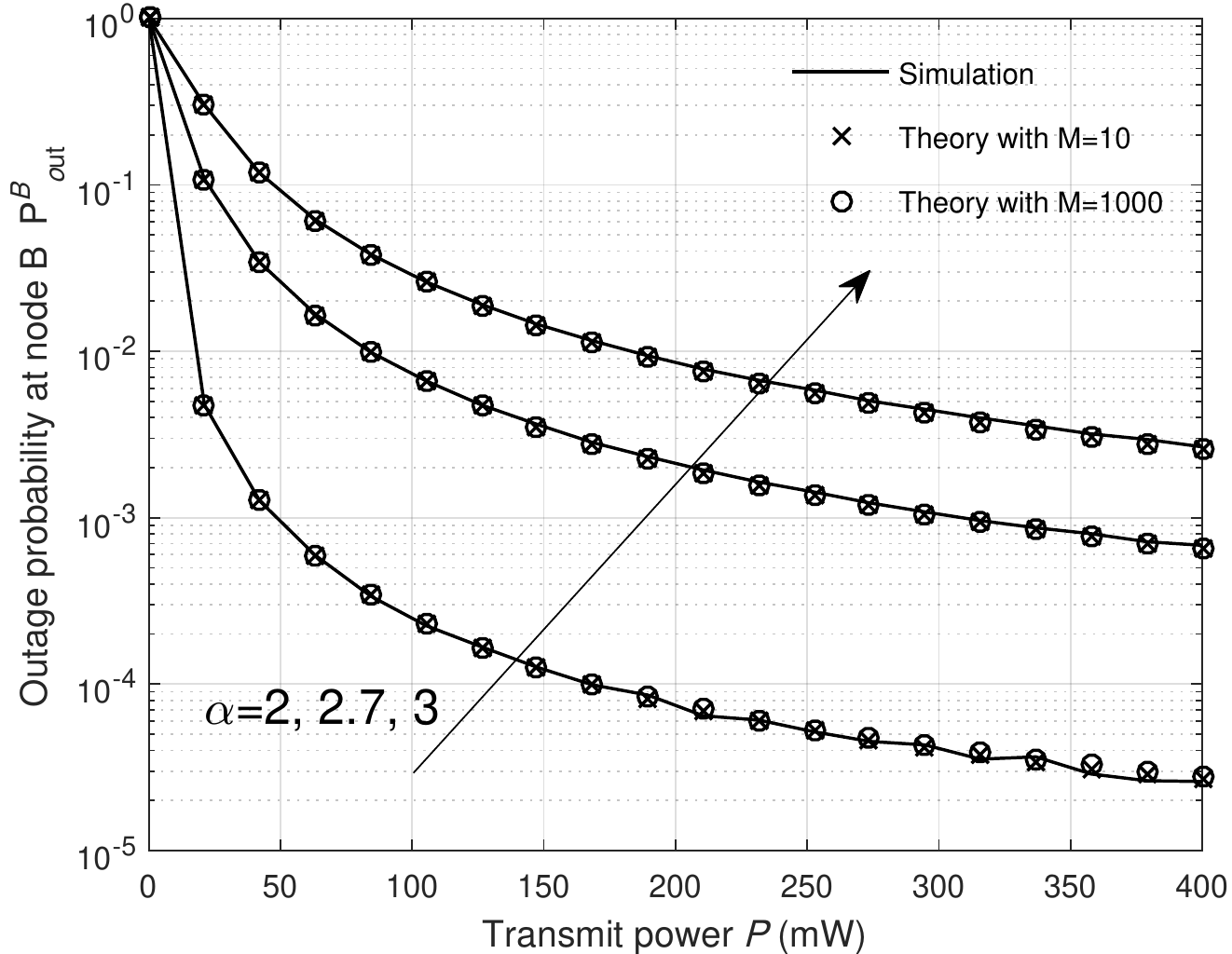}\\
  \caption{Outage probability versus transmit power $P$.}\label{fig1}
\end{figure}
\begin{figure}
  \centering
  \includegraphics[width=0.45\textwidth]{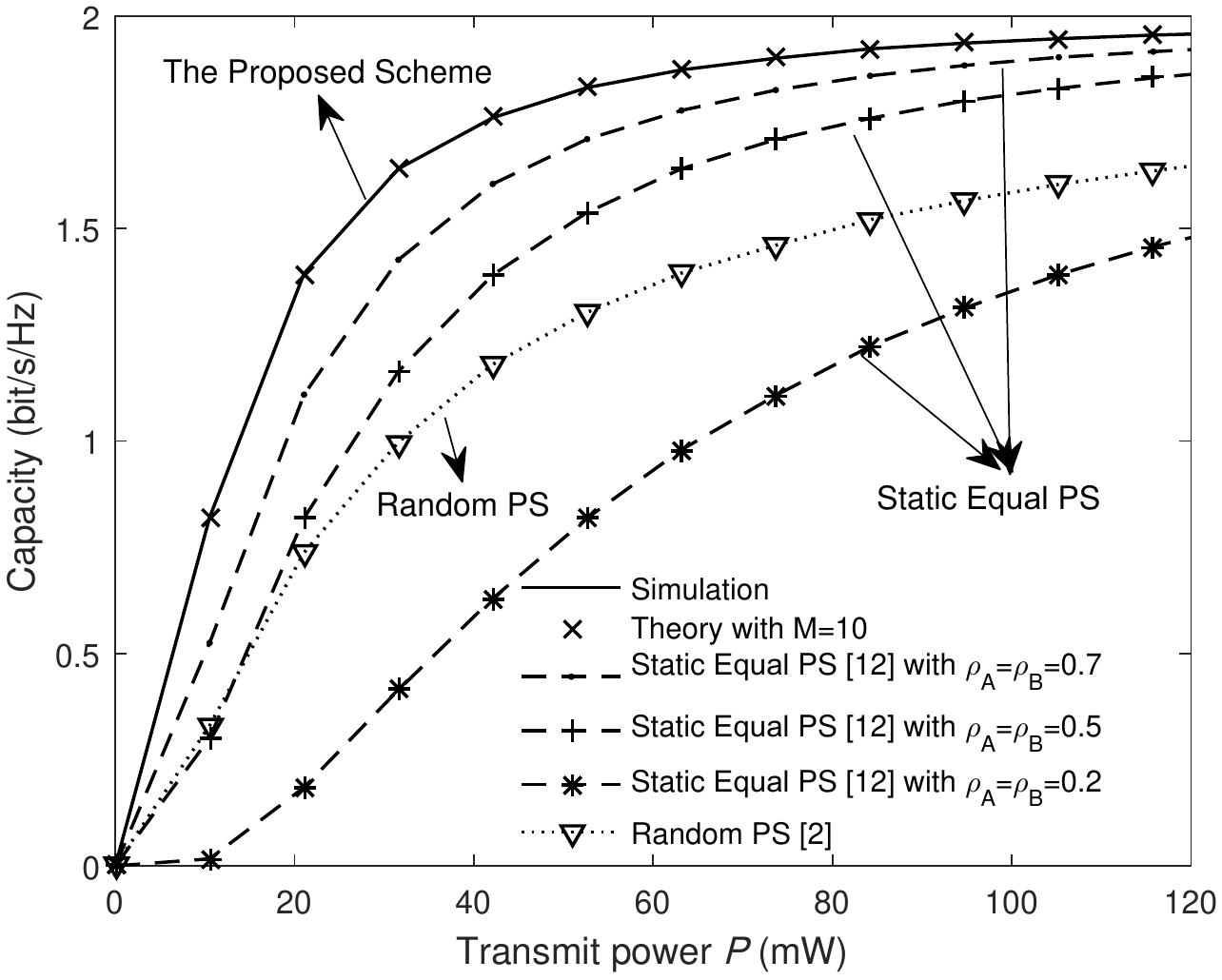}\\
  \caption{Capacity versus transmit power $P$ with $\alpha=3$.}\label{fig2}
\end{figure}

\section{Simulations}
In this section, we validate the performance of the proposed scheme and the derived outage probability via $1 \times 10^{6}$ Monte-Carlo simulations. Unless otherwise specified, the simulation parameters are set as follows: ${d_{A}} = 15\;{\rm{meters}}$, ${d_{B}} = 10\;{\rm{meters}}$\footnote{The distances for source-relay and relay-destination links reflects the state of the art
for sensor network applications. The typical single hop communications distance ranges from several meters to tens of meters.}, $\alpha=3$, $\beta=\frac{1}{3}$ and $\sigma^{2}  = -90$ dBm. The transmission rate is assumed as $U=3\;{\rm{bit/s/Hz}}$ and the corresponding SNR threshold $\gamma_{\rm{th}}$ is $2^{U}-1$. The transmit power of the source is set to be $10$ mW.
We employ the piecewise linear EH model with $N=4$, where $P_{\rm{th}}=[10,57.68,230.06,100]$ uW, $\{a_{k}\}^{3}_{1}=[0.3899,0.6967,0.1427]$, $\{b_{k}\}^{3}_{1}=[-1.6613,-19.1737,108.2778]$ uW and $P_{\rm{m}}=250$ uW.  The accuracy of this model is verified by comparing it with the experimental data in \cite{Data1}.

Figure 2 plots the outage probability at $B$ with the piecewise linear EH model achieved by the proposed scheme versus the transmit power with $\alpha=2,2.7$, and $3$, respectively.
The theoretical results with different $M$ are computed based on Eq. \eqref{R2}, Eq. \eqref{R3}, and Eq. \eqref{R5}. It can be observed that the theoretical results match perfectly with Monte Carlo simulation results, which verifies the accuracy of the theoretical results. Besides, it can also be seen that a small $M$ (e.g. $M=10$) is sufficient to provide an accurate $P_{\rm{out}}^{B}$. Another observation is that the outage probability at node $B$ converges to the error floors when the transmit power $P$ keeps increasing, which is the main difference from the outage behaviors with the linear energy harvesting model. This is due to the fact that the harvested energy from $A$ or $B$ is constrained to $P_{\rm{m}}$ when $P$ is large enough.


%
Figure 3 plots the capacity as a function of $P$ with three PS schemes: the proposed scheme, the existing static equal scheme in \cite{2017CL}, and the random PS scheme in \cite{RandomPS}. For the random PS scheme, the PS ratio follows a uniform distribution
over the closed interval $[0, 1]$. It can be observed that the capacity increases with the increase of $P$ and converges to the  maximum value when $P$ is large enough, which perfectly matches the results in Fig. 2. Another observation is that the proposed scheme has a higher capacity than the existing schemes in \cite{RandomPS} and  \cite{2017CL}. The reason is that the proposed scheme can provide more flexibility and effectively utilize the instantaneous CSI.

\begin{figure}
  \centering
  \includegraphics[width=0.45\textwidth]{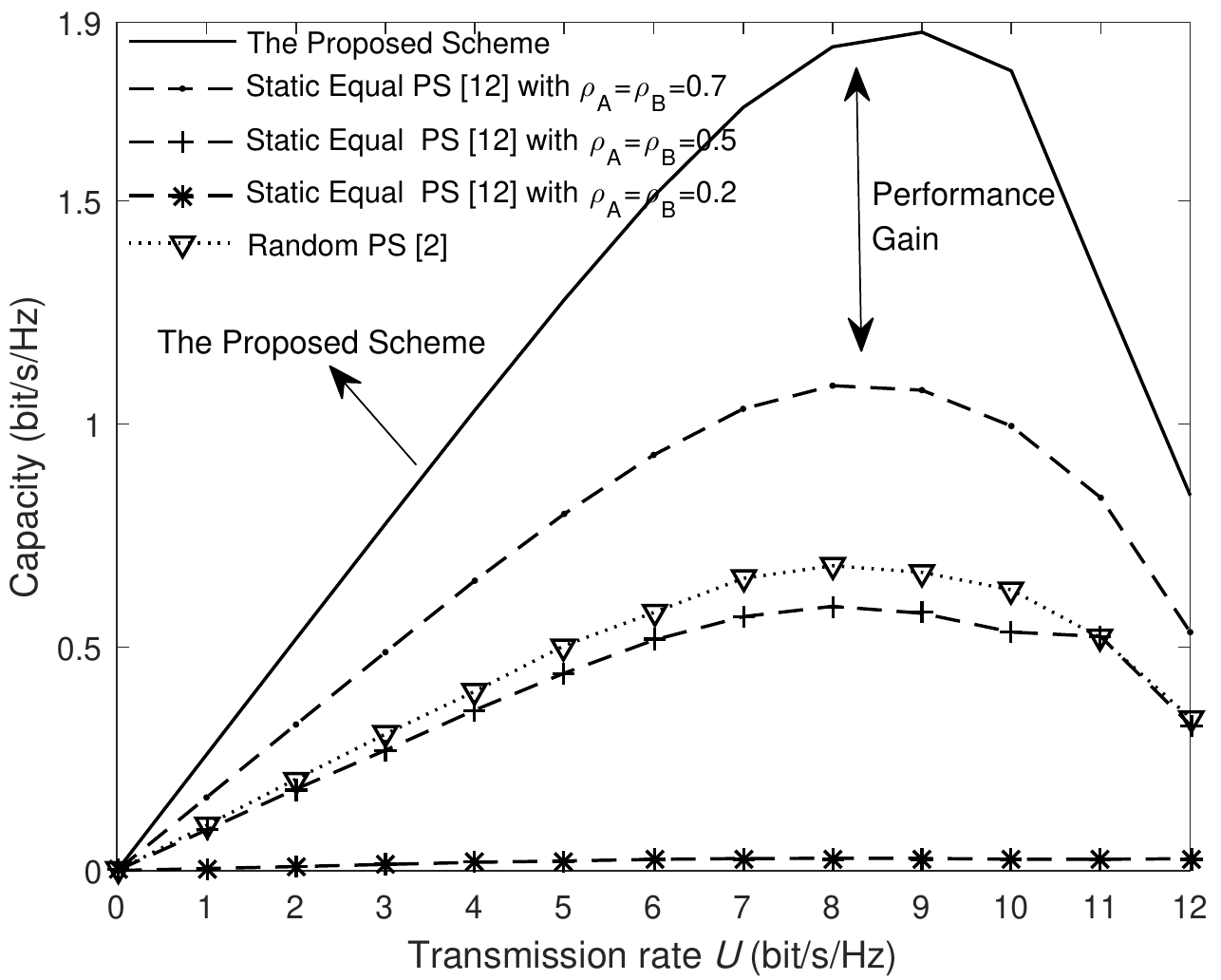}\\
  \caption{Capacity versus transmission rate $U$ with $\alpha=3$.}\label{fig3}
\end{figure}
\begin{figure}
  \centering
  \includegraphics[width=0.45\textwidth]{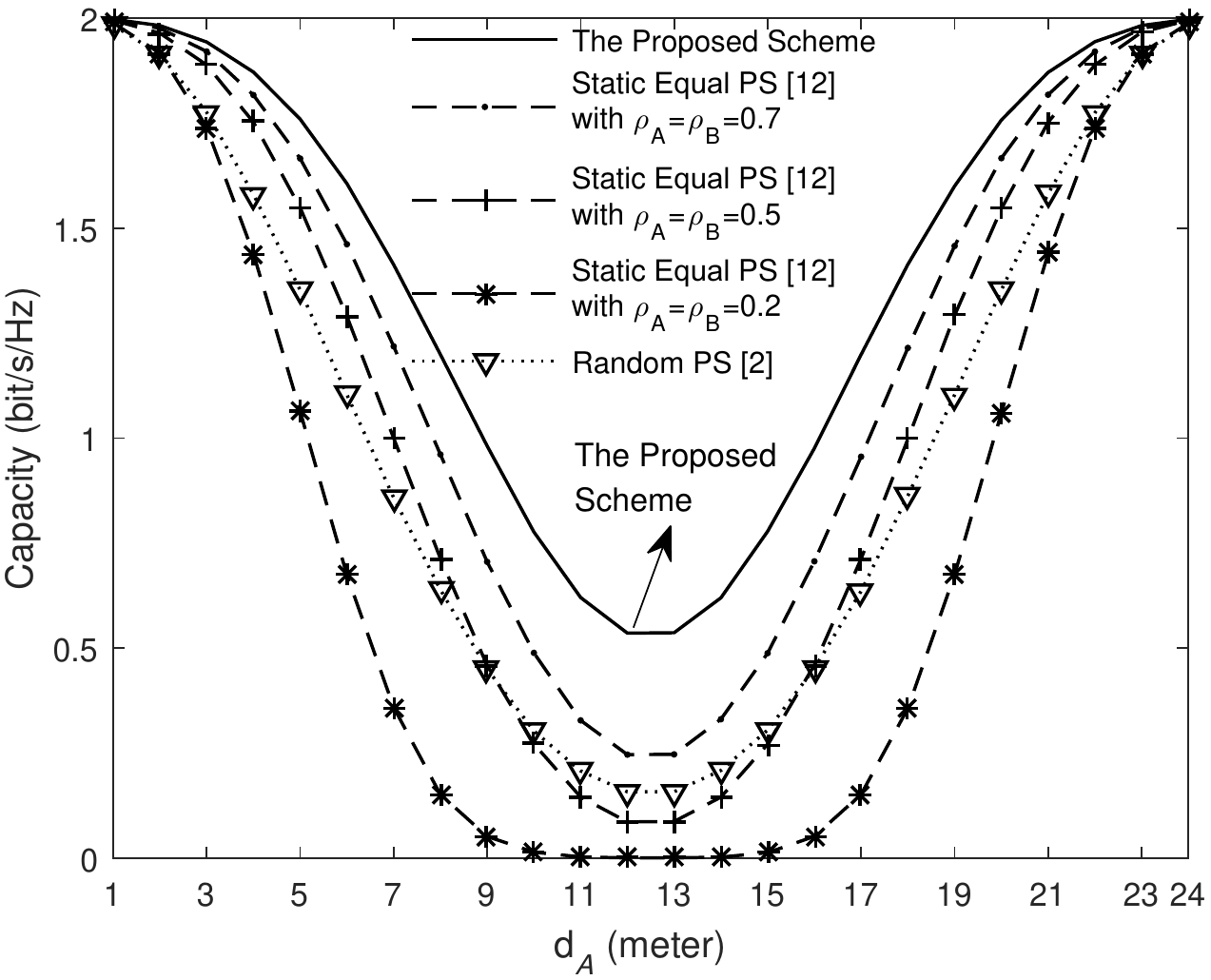}\\
  \caption{Capacity versus the distance between $A$ and $R$ $d_{A}$ with $\alpha=3$.}\label{fig4}
\end{figure}

Figure 4 plots the capacity for three PS schemes versus the transmission rate $U$. It can be seen that the capacity increases first, reaches the peak value, and then
decreases. This is because that the outage probability at $A$ or $B$ goes up with the increase of $U$ and the influence of the outage probability becomes the dominant factor to the capacity when $U$ is large enough. As shown in this figure, we can also see that the proposed scheme can provide a significant performance gain over the existing schemes.
Fig. 5 compares the capacity of various PS schemes as a function of $d_{A}$. It is assumed that $d_{A}+d_{B}=25$. Given a fixed $d_{A}$, $d_{B}$ can be computed as $25-d_{A}$. It can be observed that with the increase of $d_A$, the capacity decreases,
reaches the minimum value and then increases. This is because that the total harvested energy is higher when the relay is closer to either of the nodes.
Besides, we can see that the proposed scheme is superior to the existing schemes in terms of capacity.

\section{Conclusions}
In this paper, we have proposed a dynamic heterogeneous PS scheme to maximize the capacity of SWIPT based three-step DF TWRNs with a non-linear EH model. Specifically, by considering the heterogeneous instantaneous channel gains between the destination nodes and the relay, we have derived the closed-form expression of the optimal PS ratio for each link.
Based on the optimal PS ratios, we have derived an analytical expression for the optimal outage probability under the non-linear EH
model.
Simulation results have verified the correctness of the derived outage probability and shown that the proposed PS scheme can
achieve a higher capacity compared with the existing schemes. 

\section*{Acknowledgments}
This work was supported in part by the National Natural Science Foundation of China (Nos. 61671347, 61771368 and 61701364), Young Elite Scientists Sponsorship Program By CAST (2016QNRC001), the Fundamental Research Funds for the Central Universities, the National Natural Science Foundation of Shaanxi Province (2018JM6019)and the 111 Project of China (B08038). The work of Prof. R. Q. Hu was supported by the US National Science Foundations grants under the grants NeTS-1423348 and the EARS-1547312.

\ifCLASSOPTIONcaptionsoff
  \newpage
\fi
\bibliographystyle{IEEEtran}
\bibliography{refa}

\begin{thebibliography}{10}
\providecommand{\url}[1]{#1}
\csname url@samestyle\endcsname
\providecommand{\newblock}{\relax}
\providecommand{\bibinfo}[2]{#2}
\providecommand{\BIBentrySTDinterwordspacing}{\spaceskip=0pt\relax}
\providecommand{\BIBentryALTinterwordstretchfactor}{4}
\providecommand{\BIBentryALTinterwordspacing}{\spaceskip=\fontdimen2\font plus
\BIBentryALTinterwordstretchfactor\fontdimen3\font minus
  \fontdimen4\font\relax}
\providecommand{\BIBforeignlanguage}[2]{{%
\expandafter\ifx\csname l@#1\endcsname\relax
\typeout{** WARNING: IEEEtran.bst: No hyphenation pattern has been}%
\typeout{** loaded for the language `#1'. Using the pattern for}%
\typeout{** the default language instead.}%
\else
\language=\csname l@#1\endcsname
\fi
#2}}
\providecommand{\BIBdecl}{\relax}
\BIBdecl

\bibitem{7744827}
W.~Guo, S.~Zhou, Y.~Chen, S.~Wang, X.~Chu, and Z.~Niu, ``Simultaneous
  information and energy flow for {IoT} relay systems with crowd harvesting,''
  \emph{IEEE Commun. Mag.}, vol.~54, no.~11, pp. 143--149, November 2016.

\bibitem{RandomPS}
H.~Lee, C.~Song, S.~H. Choi, and I.~Lee, ``Outage probability analysis and
  power splitter designs for {SWIPT} relaying systems with direct link,''
  \emph{IEEE Commun. Lett.}, vol.~21, no.~3, pp. 648--651, March 2017.

\bibitem{8187650}
Z.~Chu, F.~Zhou, Z.~Zhu, R.~Q. Hu \emph{et~al.}, ``Wireless powered sensor
  networks for {I}nternet of {T}hings: Maximum throughput and optimal power
  allocation,'' \emph{IEEE Internet Things J.}, vol.~5, no.~1, pp. 310--321,
  Feb 2018.

\bibitem{8108186}
H.~Sun, Q.~Wang, S.~Ahmed, and R.~Q. Hu, ``Non-orthogonal multiple access in a
  mmwave based {IoT} wireless system with {SWIPT},'' in \emph{2017 IEEE 85th
  VTC Spring}, June 2017, pp. 1--5.

\bibitem{ShiTvt}
L.~Shi, L.~Zhao, K.~Liang, and H.~H. Chen, ``Wireless energy transfer enabled
  {D2D} in underlaying cellular networks,'' \emph{IEEE Trans. Veh. Technol.},
  vol.~67, no.~2, pp. 1845--1849, Feb 2018.

\bibitem{R1}
Z.~Ding, I.~Krikidis, B.~Sharif, and H.~V. Poor, ``Wireless information and
  power transfer in cooperative networks with spatially random relays,''
  \emph{IEEE Trans. Wireless Commun.}, vol.~13, no.~8, pp. 4440--4453, Aug
  2014.

\bibitem{T21}
A.~Alsharoa, H.~Ghazzai, A.~E. Kamal, and A.~Kadri, ``Wireless {RF}-based
  energy harvesting for two-way relaying systems,'' in \emph{Proc. IEEE WCNC},
  April 2016, pp. 1--6.

\bibitem{T23}
T.~P. Do, I.~Song, and Y.~H. Kim, ``Simultaneous wireless transfer of power and
  information in a decode-and-forward two-way relaying network,'' \emph{IEEE
  Trans. Wireless Commun.}, vol.~16, no.~3, pp. 1579--1592, March 2017.

\bibitem{T31}
Y.~Liu, L.~Wang, M.~Elkashlan, T.~Q. Duong, and A.~Nallanathan, ``Two-way
  relaying networks with wireless power transfer: Policies design and
  throughput analysis,'' in \emph{Proc. GLOBECOM}, Dec 2014, pp. 4030--4035.

\bibitem{EL}
S.~T. Shah, K.~W. Choi, S.~F. Hasan, and M.~Y. Chung, ``Energy harvesting and
  information processing in two-way multiplicative relay networks,''
  \emph{Electron. Lett.}, vol.~52, no.~9, pp. 751--753, 2016.

\bibitem{8287997}
Z.~Wang, Y.~Li, Y.~Ye, and H.~Zhang, ``Dynamic power splitting for three-step
  two-way multiplicative {AF} relay networks,'' in \emph{Proc. IEEE VTC-Fall},
  Sept 2017, pp. 1--5.

\bibitem{2017CL}
N.~T.~P. Van, S.~F. Hasan, X.~Gui, S.~Mukhopadhyay, and H.~Tran, ``Three-step
  two-way decode and forward relay with energy harvesting,'' \emph{IEEE Commun.
  Lett.}, vol.~21, no.~4, pp. 857--860, April 2017.

\bibitem{8355777}
G.~Lu, L.~Shi, and Y.~Ye, ``Maximum throughput of {TS/PS} scheme in an {AF}
  relaying network with non-linear energy harvester,'' \emph{IEEE Access},
  vol.~6, pp. 26\,617--26\,625, 2018.

\bibitem{YeICC}
Y.~Ye, Y.~Li, F.~Zhou, N.~Al-Dhahir, and H.~Zhang, ``Power splitting-based
  {SWIPT} with dual-hop {DF} relaying in the presence of a direct link,''
  \emph{IEEE Syst. J.}, pp. 1--5, 2018.

\bibitem{Ng}
E.~Boshkovska, D.~W.~K. Ng, N.~Zlatanov, and R.~Schober, ``Practical non-linear
  energy harvesting model and resource allocation for {SWIPT} systems,''
  \emph{IEEE Commun. Lett.}, vol.~19, no.~12, pp. 2082--2085, Dec 2015.

\bibitem{Data1}
T.~Le, K.~Mayaram, and T.~Fiez, ``Efficient far-field radio frequency energy
  harvesting for passively powered sensor networks,'' \emph{IEEE J. Solid State
  Circuits}, vol.~43, no.~5, pp. 1287--1302, May 2008.

\bibitem{Data2}
J.~Guo and X.~Zhu, ``An improved analytical model for {RF-DC} conversion
  efficiency in microwave rectifiers,'' in \emph{Proc. IEEE MTT S Int.
  Microwave Symp. Dig.}, June 2012, pp. 1--3.

\bibitem{Tcom}
X.~Yue, Y.~Liu, S.~Kang, A.~Nallanathan, and Y.~Chen, ``Modeling and analysis
  of two-way relay non-orthogonal multiple access systems,'' \emph{IEEE Trans.
  Commun.}, vol.~PP, no.~99, pp. 1--1, 2018.

\end{thebibliography}

\end{document}